\documentclass[floatfix,aps,superscriptaddress,showpacs,preprint]{revtex4}
\usepackage{amsmath}
\usepackage{graphicx}
\usepackage{amssymb}

\begin{document}

\title{Geometric measure of quantum discord under decoherence}

\author{Xiao-Ming Lu} \affiliation{Zhejiang Institute of Modern
  Physics, Department of Physics, Zhejiang University, Hangzhou
  310027, China}

\author{Zhengjun Xi} \affiliation{Zhejiang Institute of Modern
  Physics, Department of Physics, Zhejiang University, Hangzhou
  310027, China} \affiliation{College of Computer Science, Shaanxi
  Normal University, Xi'an 710062, China}

\author{Zhe Sun} \affiliation{Department of Physics, Hangzhou Normal
  University, Hangzhou, 310036, China}

\author{Xiaoguang Wang} \email{xgwang@zimp.zju.edu.cn}
\affiliation{Zhejiang Institute of Modern Physics, Department of
  Physics, Zhejiang University, Hangzhou 310027, China}

\begin{abstract}
  The dynamics of a geometric measure of the quantum discord (GMQD)
  under decoherence is investigated. We show that the GMQD of a
  two-qubit state can be alternatively obtained through the singular
  values of a $3\times4$ matrix whose elements are the expectation
  values of Pauli matrices of the two qubits. By using Heisenberg
  picture, the analytic results of the GMQD is obtained for three
  typical kinds of the quantum decoherence channels. We compare the
  dynamics of the GMQD with that of the quantum discord and of
  entanglement. We show that a  sudden change in the decay rate of the GMQD does not always imply
  that of the quantum discord, and vice versa. We also give a general analysis on the sudden change in behavior and find that at least for the Bell diagonal states, the sudden changes in decay rates of the GMQD and that of the quantum discord occur simultaneously.
\end{abstract}

\maketitle

\section{Introduction}

Correlations of bipartite states, including classical and quantum
parts, are of great importance and interest in quantum information
theory. Quantum discord was proposed to quantify the quantum
correlations~\cite{Ollivier2001,Henderson2001}.  It was suggested that
the quantum discord, rather than entanglement, is responsible for the
efficiency of a quantum computer, which is confirmed both
theoretically~\cite{Datta2008} and experimentally~\cite{Lanyon2008}.
A great deal of efforts has been devoted into the study of quantum
discord~\cite{Ollivier2001,Sarandy2009,Werlang2009,YXChen2010,YXChen2010_2,
  Adesso2010,Ali2010,BWang2010,Dakic2010,Datta2008,Dillenschneider2008,
  Fanchini2009,Giorda2010,Henderson2001,JSXu2010,Lanyon2008,Luo2008,
  Maziero2009,Maziero2010,Mazzola2010,Modi2010,Ferraro2009,Werlang2010,
  Maziero2010_2,Bylicka2010,Datta2010,Datta2009}. Despite this, it is not easy
to obtain analytical results of the quantum discord since the
optimization procedure involved is unreachable for arbitrary bipartite
states up to now.  Even for two-qubit systems, the analytic results
are only known for a few
cases~\cite{Luo2008,Dillenschneider2008,Sarandy2009,Ali2010,Adesso2010,Giorda2010,YXChen2010},
and a general method still lacks. To avoid this difficulty and obtain
an analytic analysis, alternative approaches are needed, among which
is the geometric measure of quantum discord (GMQD). Despite not reflected 
in the present work, another advantage
of the GMQD is that it could potentially supply a way to put various correlations on an
equal footing since the geometric measure of other kinds of correlations can be defined in the same manner but with different sets of zero-correlation states. It is remarkable that a unified view of correlations has been established through the relative entropy measure of correlations in Ref.~\cite{Modi2010}. The GMQD, similar to the geometry measure of the
entanglement~\cite{Vedral1997,TCWei2003}, is defined as the nearest
distance between the given state and the set of zero-discord states.
In the present work, we use the Hilbert-Schmidt norm as the distance
of two quantum states, because for two-qubit systems the minimization
of the Hilbert-Schmidt distance over the set of zero-discord states
was resolved analytically in Ref.~\cite{Dakic2010}. However, the
quantum discord is based on the Von Neumann entropy while the GMQD is
based on the geometric distance, their behaviors may be
different. This motivates us to consider the behaviors of the GMQD
under decoherence, and compare it with the quantum discord.

Due to the inevitable interaction with environment, the dynamics of
the quantum discord under decoherence is of great importance. It has
received some
investigations~\cite{Maziero2009,Maziero2010,Mazzola2010,Werlang2009,Fanchini2009,BWang2010},
and was experimentally investigated in an all-optical setup most
recently~\cite{JSXu2010}.  It was found that the quantum discord may
decay in an asymptotic way under Markovian
environment~\cite{Werlang2009} and vanish only at some time points
under non-Markovian environment~\cite{BWang2010,Fanchini2009}. It can
be understood by the facts that the subset of the zero-discord states
has measure zero and is nowhere dense~\cite{Ferraro2009}. While the entanglement suffers form sudden
death~\cite{Yu2004,Eberly2007,Yu2009}, because the set of separable
states occupies finite
volume~\cite{Yu2007,Zhou2010A,Zhou2010B}. Besides, in some situations,
the decay rates of the quantum discord may be
discontinuous~\cite{Maziero2009,Mazzola2010}. This is a novel
phenomena and was observed in the recent experiment~\cite{JSXu2010}.
Notice for a tripartite system $ABC$ in pure states, 
the quantum discord of $AB$ and the
entanglement of formation (EoF) of $AC$ are connected
through a monogamy relation~\cite{Koashi2004,Cen2010}, so studying the
dynamics of the quantum discord will also be helpful to the
understanding of the dynamics of EoF.

In the present work, we investigate the GMQD under decoherence
channels and get the analytical results. Under three typical quantum
decoherence channels, we will show that the GMQD is monotonically
non-decreasing with respect to the quantum discord.  Yet, the quantum
discord may keep constant while the GMQD decreases.  We will show that
in some cases the decay rates of the GMQD and of the quantum discord
may suddenly change at the same time. However, a sudden change in the
decay rate of the GMQD does not always imply a sudden change in the
decay rate of the quantum discord, and vice versa. We demonstrate each case by instances. We also give a general analysis for the sudden change in the decay rate of the GMQD and that of the quantum discord, and show that at least for the Bell diagonal states, the sudden changes in decay rates of the GMQD and that of the quantum discord occur simultaneously. 

This paper is organized as follows. In Sec.~\ref{sec:GMQD}, we give a
brief introduction of the quantum discord and the GMQD. Then we show
that the GMQD of a two-qubit state is related to the singular values
of a peculiar $3\times4$ matrix. In Sec.~\ref{sec:GMQD_channel}, we
give a general method to obtain the GMQD under quantum decoherence
channels, and get the analytic results for three typical kinds of
quantum decoherence channel.  We investigate the sudden change in the
decay rate of the GMQD, and compare it with the case of the quantum
discord. And we also give a general analysis on the disagreement of sudden change in decay rates between the GMQD and the quantum discord. Section \ref{sec:Conclusion} is the conclusion and discussion.

\section{Geometric measure of quantum discord\label{sec:GMQD}}

Given a quantum state $\rho$ in a composite Hilbert space
$\mathcal{H}=\mathcal{H}_{A}\otimes\mathcal{H}_{B}$, the total amount
of correlation is quantified by quantum mutual
information~\cite{Groisman2005}
\begin{equation}
  \mathcal{I}(\rho)=H(\rho_{A})+H(\rho_{B})-H(\rho),
\end{equation}
where $H(\rho)\equiv-\mathrm{Tr}\left[\rho\log_{2}\rho\right]$ is the
von Neumann entropy and $\rho_{A(B)}=\mathrm{Tr}_{B(A)}\rho$ is the
reduced density matrix by tracing out system $B(A)$. If we take the
system $A$ as the apparatus, the quantum discord is defined as
follows~\cite{Ollivier2001,Henderson2001}
\begin{equation}
  \mathcal{D}_{A}(\rho)=\mathcal{I}(\rho)-\mathcal{C}_{A}(\rho),\label{eq:quantum_discord}
\end{equation}
which is the difference of the total amount of correlation
$\mathcal{I}(\rho)$ and the classical correlation
$\mathcal{C}_{A}(\rho)$. Here the classical correlation is defined
by \begin{equation}
  \mathcal{C}_A(\rho)=\max_{\{E_{k}\}}\mathcal{I}(\rho|\{E_{k}\}),\label{eq:classical_correlation}\end{equation}
where $\mathcal{I}(\rho|\{E_{k}\})$ is a variant of quantum mutual
information based on a given measurement basis $\{E_{k}\}$ on system
$A$ as follows
\begin{equation}
  \mathcal{I}(\rho|\{E_{k}\})=H(\rho_{B})-\sum_{k}p_{k}H(\rho_{B|k}).\label{eq:variant_mutual_information}
\end{equation}
$\rho_{B|k}=\mathrm{Tr}_{A}[(E_{k}\otimes\openone)\rho]/p_{k}$ is the
postmeasurement state of $B$ after obtaining outcome $k$ on $A$ with
the probability $p_{k}=\mathrm{Tr}[(E_{k}\otimes\openone)\rho]$.
$\{E_{k}\}$ is a set of one-dimensional projectors on
$\mathcal{H}_{A}$, and $\openone$ is the $2\times2$ identity operator.

In Ref.~\cite{Dakic2010}, Daki\'{c} \emph{et al.} proposed a geometric
measure of quantum discord defined by
\begin{equation}
  D_{A}^{g}(\rho):=\min_{\chi\in\Omega_{0}}||\rho-\chi||^{2},
\end{equation}
where $\Omega_{o}$ denotes the set of zero-discord states and
$||X||^{2}:=\mathrm{Tr}(X^{\dagger}X)$ is the Hilbert-Schmidt
norm. The subscript $A$ of $D_{A}^g$ implies that the measurement is
taken on the system $A$. For two-qubit systems, a zero-discord state
is of the form
$\chi=p_{1}|\psi_{1}\rangle\langle\psi_{1}|\otimes\rho_{1}+p_{2}|\psi_{2}\rangle\langle\psi_{2}|\otimes\rho_{2}$
with $|\psi_1\rangle$ and $|\psi_2\rangle$ two arbitrary orthogonal
states. And a general state can be written in Bloch representation
\cite{Schlienz1995}:
\begin{eqnarray}
  \rho & = & \frac{1}{4}\left[\openone\otimes\openone+\sum_{i}^{3}(x_{i}\sigma_{i}\otimes\openone+y_{i}\openone\otimes\sigma_{i})+\sum_{i,j=1}^{3}R_{ij}\sigma_{i}\otimes\sigma_{j}\right]
\end{eqnarray}
with $x_{i}$, $y_{i}$, and $R_{ij}$ real parameters, and
$\sigma_{i=1,2,3}$ Pauli matrices. Then an explicit expression of the
GMQD is obtained as~\cite{Dakic2010}:
\begin{equation}
  D_{A}^{g}(\rho)=\frac{1}{4}\left(||x||^{2}+||R||^{2}-k_{\mathrm{max}}\right),\label{eq:GMQD_original}
\end{equation}
where $x=(x_{1},x_{2},x_{3})^{T}$, $R$ is the matrix with elements
$R_{ij}$, and $k_{\mathrm{max}}$ is the largest eigenvalue of matrix
$K=xx^{T}+RR^{T}$.

Now, we introduce an alternative form which will be convenient when we
consider the evolution of the GMQD under decoherence. First, we
introduce a matric $\mathcal{R}$ defined by
\begin{equation}
  \mathcal{R}=\left[\begin{array}{cc}
      1 & y^{T}\\
      x & R\end{array}\right],
\end{equation}
and another $3\times4$ matric $\mathcal{R}^{\prime}$ obtained through
deleting the first row of $R$, i.e., $\mathcal{R}^{\prime}=(x,R)$.
Here $\mathcal{R}$ is just the expectation matric with the elements
$\mathcal{R}_{ij}=\mathrm{Tr}[\rho\sigma_{i}\otimes\sigma_{j}]$ for
$i,j=0,1,2,3$, and $\sigma_{0}=\openone$ is defined. The definition of
$\mathcal{R}^{\prime}$ leads to
$K=\mathcal{R}^{\prime}(\mathcal{R}^{\prime})^{T}$.  After singular
value decomposition, we have $\mathcal{R}^\prime=U \Lambda V^T$, where
$U$ and $V$ are $3\times3$ and $4\times4$ orthogonal matrices, and
$\Lambda$ has only diagonal elements $\Lambda_{ij}=\lambda_i
\delta_{ij}$ with $\lambda_i$ the so-called singular values of the
matrix $\mathcal{R}^\prime$.  Then the eigenvalues of the matrix $K$
can be expressed as $\lambda_i^2$. Considering
$||x||^{2}+||R||^{2}=\mathrm{Tr}K$, we get an alternative compact form
of $D_{A}^{g}(\rho)$:
\begin{equation}
  D_{A}^{g}(\rho)=\frac{1}{4}\left[\left(\sum_{k}\lambda_{k}^{2}\right)-\max_{k}\lambda_{k}^{2}\right],\label{eq:GMQD}
\end{equation}
where the summation and maximization are taken over all the non-zero
singular values $\lambda_k$ of $\mathcal{R}^\prime$.  This alternative
form will be convenient when we consider the evolution of the GMQD
under decoherence.

\section{Geometric measure of quantum discord under quantum
  decoherence channels\label{sec:GMQD_channel}}

A quantum channel can be described in the Kraus
representation 
\begin{equation}
  \mathcal{E}(\rho)=\sum_{\mu}K_{\mu}\rho
  K_{\mu}^{\dagger},
\end{equation} 
where $K_{\mu}$ are Kraus operators satisfying $\sum_{\mu}K_{\mu}^{\dagger}K_{\mu}=\openone$.  As we
discussed in the previous section, to obtain the GMQD, we need to know
the expectation values of the Pauli matrices of the two qubits for the
state $\mathcal{E}(\rho)$. So we turn to the Heisenberg picture to
describe quantum channels via the map~\cite{Wang2010}
\begin{equation}
  \mathcal{E}^{\dagger}(A)=\sum_{\mu}K_{\mu}^{\dagger}AK_{\mu}
\end{equation}
with $A$ an arbitrary observable. Then the expectation value of $A$
can be obtained through $\langle
A\rangle=\mathrm{Tr}\left[A\mathcal{E}(\rho)\right]=\mathrm{Tr}\left[\mathcal{E}^{\dagger}(A)\rho\right]$.
Because an arbitrary Hermitian operator on $\mathbb{C}^{2}$ can be
expressed by $A=\sum_{i=0}^{3}r_{i}\sigma_{i}$ with
$r_{i}\in\mathbb{R}$, then a quantum channel for a qubit can be
characterized by the transmission matrix $M$ defined through
\begin{equation}
  \mathcal{E}^{\dagger}(\sigma_{i})=\sum_{j}M_{ij}\sigma_{j}\quad\mathrm{or\quad}M_{ij}=\frac{1}{2}\mathrm{Tr}\left[\mathcal{E}^{\dagger}(\sigma_{i})\sigma_{j}\right].
  \label{eq:Transmission_Matrix}
\end{equation}
Since $\mathrm{Tr}[\mathcal{E}^{\dagger}(\sigma_{i})\rho]=\sum_j
M_{ij}\mathrm{Tr}[\sigma_{j}\rho]$, $M_{ij}$ actually describes the
transformation of the polarized vector
$P_{i}\equiv\mathrm{Tr}[\sigma_{j}\rho]$.

Now we consider the case of two qubits under local decoherence
channels, i.e.,
$\rho=[\mathcal{E}_{A}\otimes\mathcal{E}_{B}](\rho_{0})$. To obtain
the GMQD of the output state $\rho$ through the channel, we need to
get the expectation matrix $\mathcal{R}$. With the Heisenberg picture,
we have
\begin{equation}
  \mathcal{R}_{ij}  = \mathrm{Tr}(\mathcal{E}_{A}^{\dagger}(\sigma_{i})\otimes\mathcal{E}_{B}^{\dagger}(\sigma_{j})\rho_{0})
  =(M_{A}\mathcal{R}_{0}M_{B}^{T})_{ij},
\end{equation}
where $\mathcal{R}_{0}$ is the expectation matrix under $\rho_{0}$,
i.e.,
$(\mathcal{R}_{0})_{ij}=\mathrm{Tr}(\sigma_{i}\otimes\sigma_{j}\rho_{0})$,
and $M_{A(B)}$ is the transformation matrix characterizing the quantum
channel $\mathcal{E}_{A(B)}$. So we obtain
$\mathcal{R}=M_{A}\mathcal{R}_{0}M_{B}^{T}$.

For simplicity, we assume $\mathcal{E}^{A}$ and $\mathcal{E}^{B}$ be
identical, hereafter. Next, we consider three typical kinds of
decoherence channels: the amplitude damping channel (ADC), the phase
damping channel (PDC), and the depolarizing channel (DPC). They are
described by the set of Kraus operators
respectively~\cite{PreskillLect,NielsenBook}:
\begin{eqnarray}
  K^{\mathrm{ADC}} & = & \left\{ \sqrt{s}|0\rangle\langle0|+|1\rangle\langle1|,\ \sqrt{p}|1\rangle\langle0|\right\} ,
  \label{eq:Kraus_ADC}\\
  K^{\mathrm{PDC}} & = & \left\{ \sqrt{s}\openone,\ \sqrt{p}|0\rangle\langle0|,\ \sqrt{p}|1\rangle\langle1|\right\} ,
  \label{eq:Kraus_PDC}\\
  K^{\mathrm{DPC}} & = & \{\frac{1}{2}\sqrt{1+3s}\,\openone,\ \frac{1}{2}\sqrt{p}\,\sigma_{x},\frac{1}{2}\sqrt{p}\,\sigma_{y},\ \frac{1}{2}\sqrt{p}\,\sigma_{z}\},
  \label{eq:Kraus_DPC}
\end{eqnarray}
with $s\equiv 1-p$. Here the real parameter $p\in[0,1]$ may be
time-dependent in some realistic
setup~\cite{NielsenBook,PreskillLect}. For instance, for the PDC, the
parameter $s$ may be like $\exp(-\gamma t)$ with $\gamma$ the rate of
damping.

From Eqs.~(\ref{eq:Transmission_Matrix}), (\ref{eq:Kraus_ADC}),
(\ref{eq:Kraus_PDC}), and (\ref{eq:Kraus_DPC}), the transmission
matrix $M$ of each channel can be got through the transformation of
the Pauli matrices in the Heisenberg picture~\cite{Wang2010} as
\begin{equation}
  M_{\mathrm{ADC}}  =  \left[\begin{array}{cccc}
      1 & 0 & 0 & 0\\
      0 & \sqrt{s} & 0 & 0\\
      0 & 0 & \sqrt{s} & 0\\
      -p & 0 & 0 & s\end{array}\right],\quad
  M_{\mathrm{PDC}}  =  \left[\begin{array}{cccc}
      1 & 0 & 0 & 0\\
      0 & s & 0 & 0\\
      0 & 0 & s & 0\\
      0 & 0 & 0 & 1\end{array}\right],\quad
  M_{\mathrm{DPC}}  =  \left[\begin{array}{cccc}
      1 & 0 & 0 & 0\\
      0 & s & 0 & 0\\
      0 & 0 & s & 0\\
      0 & 0 & 0 & s\end{array}\right].
\end{equation}
For simplicity, here we first take as the input states of two-qubit
system the Bell diagonal states~\cite{Luo2008,Mazzola2010}
\begin{equation}
  \rho=\frac{1}{4}\left(\openone\otimes\openone+\sum_{i=1}^{3}c_{i}\sigma_{i}\otimes\sigma_{i}\right),\label{eq:3c_states}
\end{equation}
which includes the Werner states $(|c_{1}|=|c_{2}|=|c_{3}|=c)$ and
Bell states $(|c_{1}|=|c_{2}|=|c_{3}|=1)$. This state is physical if
the vector $(c_{1},c_{2},c_{3})$ belongs to the tetrahedron defined by
the set of the vertices $(-1,-1,-1)$, $(-1,1,1)$, $(1,-1,1)$ and
$(1,1,-1)$~\cite{Horodecki1996}. This restriction can be described by
the following conditions~\cite{Horodecki1996,Luo2008}:
\begin{eqnarray}
  \sum_{i=1}^{3}c_{i} & \in & [-3,1],\nonumber \\
  c_{i}-c_{j}-c_{k} & \in & [-3,1]\ \mbox{for }i\neq j\neq k.\label{eq:physical_restriction}
\end{eqnarray}

For states~(\ref{eq:3c_states}),
$\mathcal{R}_{0}=\mathrm{diag}\{1,c_{1},c_{2},c_{3}\}$ is of diagonal
form. From the relation $\mathcal{R}=M\mathcal{R}_{0}M^{T}$, we get
$\mathcal{R}$ under the ADC, the PDC, the DPC respectively:
\begin{eqnarray}
  \mathcal{R}_{\mathrm{ADC}} & = & \left[\begin{array}{cccc}
      1 & 0 & 0 & -p\\
      0 & c_{1}s & 0 & 0\\
      0 & 0 & c_{2}s & 0\\
      -p & 0 & 0 & c_{3}s^{2}+p^{2}\end{array}\right],\label{eq:R_ADC}\\
  \mathcal{R}_{\mathrm{PDC}}&=&\mathrm{diag}\left\{1,c_1s^2,c_2s^2,c_3\right\},\\
  \mathcal{R}_{\mathrm{DPC}}&=&\mathrm{diag}\left\{1,c_1s^2,c_2s^2,c_3s^2\right\}.
\end{eqnarray}
$\mathcal{R}^\prime$ is obtained by deleting the first row of the
matrix $\mathcal{R}$, for ADC, PDC, DPC respectively.  Calculating the
singular values of each $\mathcal{R}^{\prime}$ for these three
decoherence channels, and substituting them into Eq.~(\ref{eq:GMQD}),
we finally obtain the GMQD as follows
\begin{align}
  D_{\mathrm{ADC}}^{g} &=\frac{1}{4}\left[s^{2}\left(c_{1}^{2}+c_{2}^{2}\right)+p^{2}+(p^{2}+c_{3}s^{2})^{2}-\max\left\{ \left(sc_{1}\right)^{2},\left(sc_{2}\right)^{2},p^{2}+(p^{2}+c_{3}s^{2})^{2}\right\} \right],\label{eq:GMQD_ADC}\\
  D_{\mathrm{PDC}}^{g} & =  \frac{1}{4}\left[s^{4}(c_{1}^{2}+c_{2}^{2})+c_{3}^{2}-\max\left\{ \left(s^{2}c_{1}\right)^{2},\left(s^{2}c_{2}\right)^{2},c_{3}^{2}\right\} \right],\\
  D_{\mathrm{DPC}}^{g} & =
  \frac{1}{4}\left[s^{4}(c_{1}^{2}+c_{2}^{2}+c_{3}^{2})-\max\left\{
      \left(s^{2}c_{1}\right)^{2},\left(s^{2}c_{2}\right)^{2},\left(s^{2}c_{3}\right)^{2}\right\}
  \right].
\end{align}

For comparison with the dynamics of entanglement, we use the
concurrence defined as~\cite{Hill1997,Wootters1998}
\begin{equation}
  C=\max(0,\lambda_{1}-\lambda_{2}-\lambda_{3}-\lambda_{4}),
\end{equation}
where $\lambda_{i}$ are the square roots of the eigenvalues in
descending order of the matrix product
$\varrho=\rho\left(\sigma_{y}\otimes\sigma_{y}\right)\rho^{*}\left(\sigma_{y}\otimes\sigma_{y}\right)$
with $\rho^*$ the complex conjugate of the two-qubit density matrix
$\rho$.

Now we assume $p(t)$ is a smooth function about time $t$, so we
investigate the evolution of the correlation quantities (the GMQD,
quantum discord and concurrence) along with $p(t)$ instead of $t$. To
investigate the behaviors of the GMQD, the quantum discord and the
concurrence, we consider some examples. The first one is the initial
state $(c_{1}=1,\ c_{2}=-c_{3},\ c_{3}=0.6)$ \cite{Mazzola2010} under
the PDC. For this case, we obtain
\begin{eqnarray}
  D_{\mathrm{PDC}}^{g}(p) & = & \min\left\{ D_{1}^{g}(p),D_{2}^{g}(p)\right\} ,\nonumber \\
  D_{1}^{g}(p) =  \frac{17}{50}(1-p)^{4},&&
  D_{2}^{g}(p) =  \frac{9}{100}\left[1+(1-p)^{4}\right].
\end{eqnarray}

\begin{figure}[!htb]
  \begin{centering}
    \includegraphics[scale=0.4]{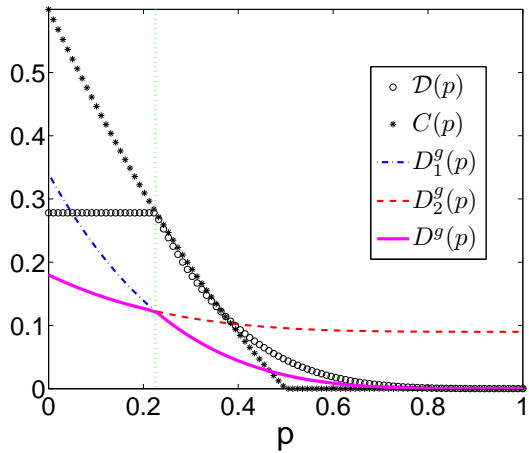}
    \par\end{centering}
  \caption{\label{fig:PDC} (Color online) Discord $D(p)$, concurrence
    $C(p)$ and the geometry measure of discord $D^{g}(p)$ under PDC,
    as functions of $p$, for the input state taken as $c_{1}=1$,
    $c_{2}=-c_{3}$ and $c_{3}=0.6$.}
\end{figure}

In Fig.~\ref{fig:PDC}, we show that the GMQD is monotonically
non-decreasing with respect to the quantum discord. When $p\leq
1-\sqrt{3/5}$, the GMQD decreases while the quantum discord keeps
constant. It is remarkable that regime where the quantum discord is
unaffected by the noisy environment is important for the
implementation of a quantum computer~\cite{Mazzola2010}.  The
entanglement may disappear completely after a finite time, known as
entanglement-sudden-death (ESD)~\cite{Yu2004,Eberly2007,Yu2009}.  In
cases where ESD occurs, contrarily, the quantum discord is more robust
than the concurrence~\cite{Werlang2009}, so does the GMQD, see
Fig.~\ref{fig:PDC}. The discontinuity of the decay rates occurs at
$p=1-\sqrt{3/5}$ (see the dotted line in Fig.~\ref{fig:PDC}). For the
GMQD, this kind of sudden change occurs when the maximum singular
value of the matrix $\mathcal{R}^{\prime}$ jumps from one family to
another one, see $D_{1}^{g}(p)$ and $D_{2}^{g}(p)$ in
Fig.~\ref{fig:PDC}.

\begin{figure}[tbh]
  \includegraphics[scale=0.4]{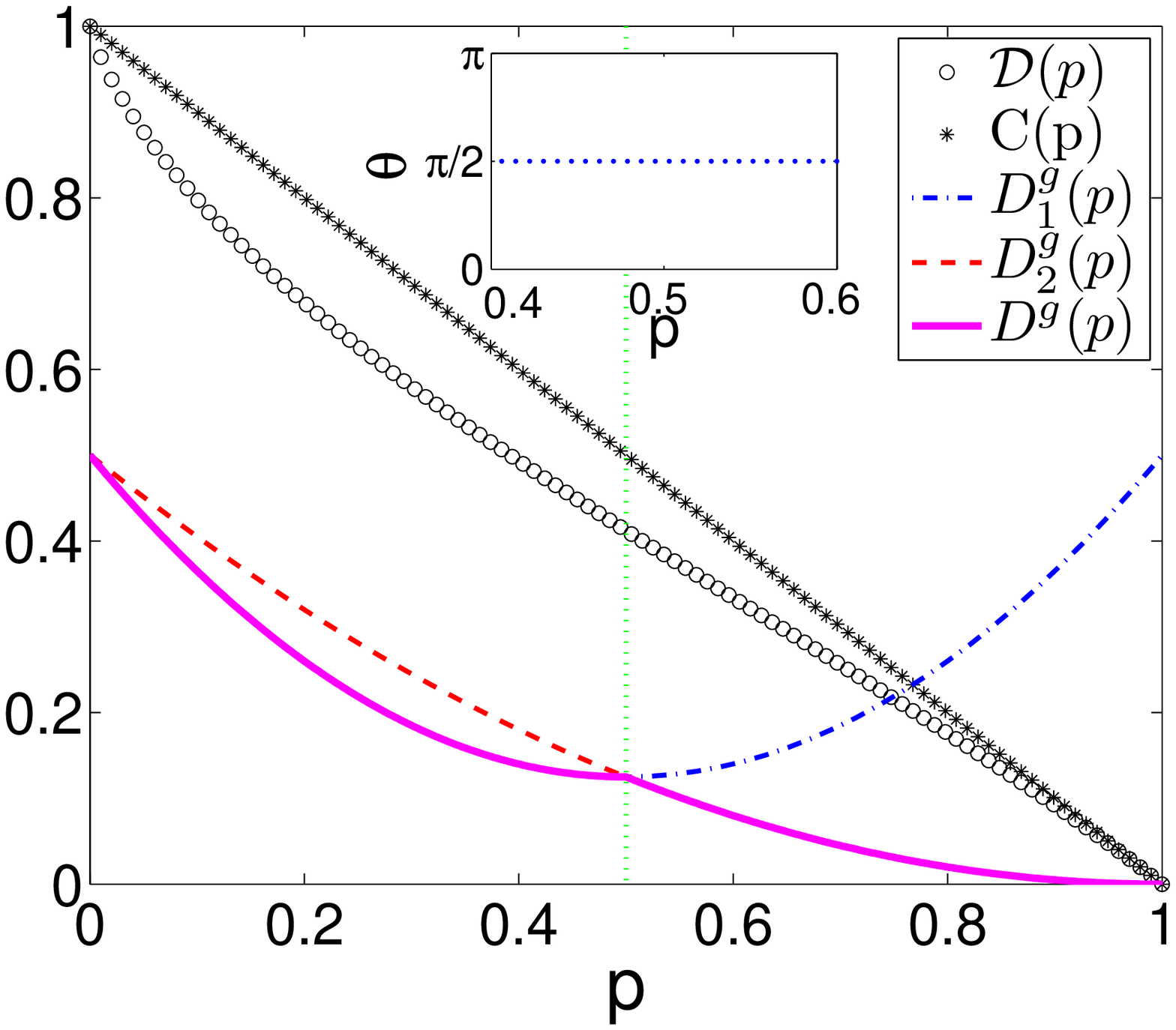}
  \includegraphics[scale=0.4]{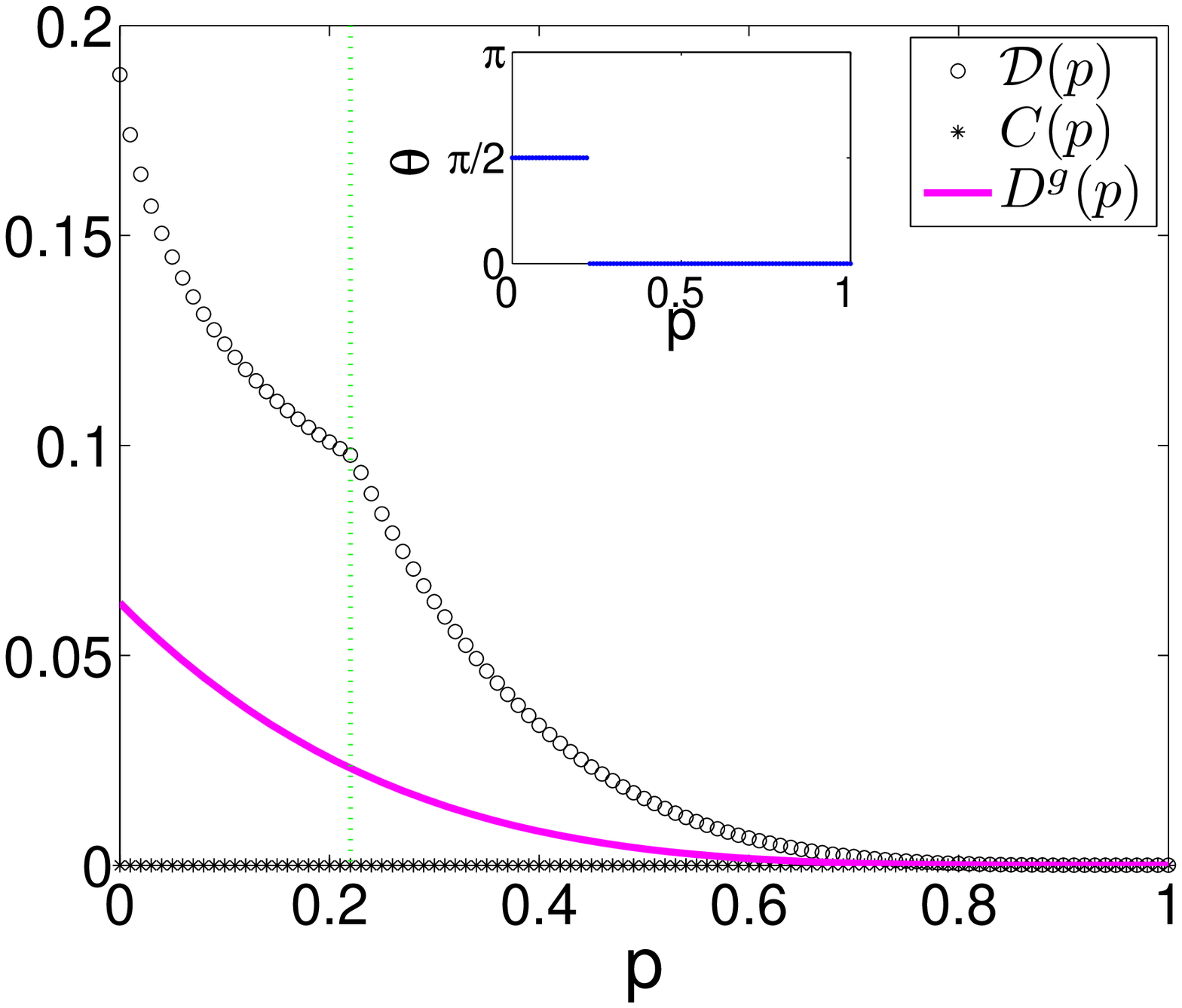}
  \caption{\label{fig:instances} (Color online) Discord $D(p)$, concurrence
    $C(p)$ and the geometry measure of discord $D^{g}(p)$ as functions
    of $p$. The left subfigure is plotted for the case of the Bell state
    ($c_{1}=c_{2}=c_{3}=-1$) under the ADC. The right subfigure is
    plotted for the case of the states (\ref{eq:third_example}) with
    $c_1=0.5$, $c_2=0$, $c_3=0.5$ and $d=-0.5$ under the PDC. In the
    inset of each figure, we plot the values of $\theta$ which
    maximize $\mathcal{I}(\rho(p)|\{E_{k}(\theta,\phi)\})$ at
    different $p$, where $E_{k}(\theta,\phi)=(\openone+\vec{n}\cdot\vec{\sigma})/2$ with $n=(\sin\theta\cos\phi,\sin\theta\sin\phi,\cos\theta)^T$.}
\end{figure}

However, the sudden change in decay rates of the GMQD may not imply
that of the quantum discord, and vice versa.  For instance of the
former, we consider the second example, a Bell state
($c_{1}=c_{2}=c_{3}=-1$) under the ADC. Substituting $c_{i}=-1$
($i=1,2,3$) into Eq.~(\ref{eq:GMQD_ADC}), we obtain
\begin{eqnarray}
  D_{\mathrm{ADC}}^g(p) & = & \min\left\{ D_{1}^{g}(p),D_{2}^{g}(p)\right\} ,\nonumber \\
  D_{1}^{g}(p) =  \frac{1}{2}(1-3p+3p^{2}),&&D_{2}^{g}(p) =  \frac{1}{2}(1-p)^{2}.
\end{eqnarray}
From the left subfigure of Fig.~\ref{fig:instances}, we can see that the sudden change in the
decay rates of the GMQD occurs at $p=0.5$, where the quantum discord
obtained numerically does not display any discontinuity in the first
derivative. To demonstrate that the sudden change is decay rates of the
quantum discord may not imply the GMQD, we consider the third
example---the input state
\begin{equation}
  \rho=\frac{1}{4}\left(\openone\otimes\openone+\sum_{i=1}^{3}c_{i}\sigma_{i}\otimes\sigma_{i}
    +d\sigma_3\otimes\openone+d\openone\otimes\sigma_3\right)
  \label{eq:third_example}
\end{equation}
under the PDC. Here the parameters are choose as $c_1=0.5$, $c_2=0$,
$c_3=0.5$ and $d=-0.5$. After some algebras similar to the case where the input states are Bell diagonal states, the GMQD is obtained as $D^g_{\mathrm{PDC}}=(1-p)^4/16$. From the right subfigure of Fig.~\ref{fig:instances}, we can see that the sudden change in decay rates of the quantum discord occurs at $p\simeq 0.22$, where the GMQD does not display any discontinuity in the first derivative.

In the following, we give a general analysis on the sudden change in decay rates of the GMQD and the quantum discord. Hereafter, we assume that the elements of the two-qubit density matrix are smooth functions of $p$, then the sudden changes in decay rate of both the GMQD and the quantum discord are induced by the optimization procedure involved in their definitions. For the GMQD, the optimization procedure is to find the closest one $\tilde \chi$ among the zero-discord state $\chi=p_1\Pi_1\otimes\rho_1+p_2\Pi_2\otimes\rho_2$. Here $\Pi_1$ and $\Pi_2$ are orthogonal projective operators and can be expressed by $\Pi_1=(\openone+\sum_{i=1}^3e_i\sigma_i)/2$ and $\Pi_2=\openone-\Pi_1$ with $e_i$ the components of  a unit vector $e=(e_1,e_2,e_3)^T$ on the Bloch sphere. In Ref.~\cite{Dakic2010}, Daki\'{c} \emph{et al.} obtain the results (\ref{eq:GMQD_original}) where $k_\mathrm{max}$ can be expressed by
\begin{equation}
  k_{\mathrm{max}}(p)=\max_{|e|=1} [e^T K(p) e].
\end{equation}
So the sharp features of the GMQD are caused by the optimization over $e$. Then the closest zero-discord state is given by $\tilde\chi=\tilde p_1\tilde\Pi_1\otimes\tilde\rho_1+\tilde p_2\tilde\Pi_2\otimes\tilde\rho_2$ where $\tilde\Pi_1(p)=(\openone+\sum_{i=1}^3\tilde e_i(p)\sigma_i)/2$ with $\tilde e(p)$ the eigenvector of $K(p)$ with the largest eigenvalue. We do not care about the other parameters in the $\tilde\chi$ since they have nothing to do with the GMQD. Hence, the sudden change in decay rates of the GMQD corresponds to the sudden change of $\tilde e(p)$. On the other hand, for the quantum discord, the optimization procedure is to find an optimal projective measurement $\{\tilde{E}_k\}$ to access the classical correlation over the set of projective measurement $\{E_{k}\}$. Then the sudden change in the decay rates of the quantum discord is induced by the sudden change of $\{\tilde{E}_{k}(p)\}$ with respect to $p$. In a similar way to $\tilde\Pi_i$, the optimal projective operator can be represented by $\tilde E_1(p)=(\openone+\sum_{i=1}^3\tilde n_i \sigma_i)/2$ and $\tilde E_2=\openone - \tilde E_1$ with $n_i$ the components of a unit vector $\tilde n=(\tilde n_1,\tilde n_2,\tilde n_3)^T$ on the Bloch sphere, or equivalently $\tilde n=(\sin\theta\cos\phi,\sin\theta\sin\phi,\cos\theta)^T$. In the inset of Fig.~\ref{fig:instances}, we plot $\theta$ for the optimized measurement $\tilde E_1(\theta,\phi)$ for different point $p$. For the second example (a Bell state under the ADC), the output state are invariant under a rotation $R_{z}(\varphi)=\exp(i\varphi\sigma_{z}^{A}/2)\otimes \exp(i\varphi\sigma_{z}^{B}/2)$ and  $\mathcal{I}(\rho|\{E_k\})$ are invariant under local unitary transformation, so if $\{\tilde E_k(\theta,\phi)\}$ is the optimal measurement, then $\{\exp(i\varphi\sigma_z^A/2)E_k(\theta,\phi)\exp(-i\varphi\sigma_z^A/2)\}=\{E(\theta,\phi-\varphi)\}$
is also the optimal measurement. In other words, the optimization procedure is only relevant to $\theta$. In the inset of the left part of Fig.~\ref{fig:instances}, we can see that there is no sudden change of $\theta$. For the third example, in the inset of the right part of Fig.~\ref{fig:instances}, we can see that a sudden change of $\theta$ occurs at $p\simeq0.22$, where the decay rate of the quantum discord is discontinuous. And we do not care about $\phi$ because the sudden change of the optimal measurement has already been reflected through the discontinuity of $\theta$.

Notice that the optimal $\tilde E_k(p)$ and $\tilde \Pi_k(p) $ are both projective operators. For general states, $\{\tilde E_k(p)\}$ is not the same as $\{\tilde \Pi_k(p)\}$, which is reflected in the disagreement of their individual sudden change in decay rate. However, at least for the Bell diagonal states, $\{\tilde E_k(p)\}$ is the same as $\{\tilde\Pi_k(p)\}$, or equivalently $\tilde n(p)=\tilde e(p)$. This can be seen from the following analysis. In Ref.~\cite{Luo2008}, Luo solved the optimization analytically for the Bell diagonal states, and $\tilde n$ is found to be the eigenvector of $RR^T$ with the largest eigenvalue. On the other hand, in Ref.~\cite{Dakic2010}, Daki\'{c} \emph{et al.} showed that $\tilde{e}$ is the eigenvector of $K=xx^T+RR^T$ with the largest eigenvalue. For the Bell states $K=RR^T$ due to $x=0$. Hence, we get $\tilde n(p)=\tilde e(p)$. So it is concluded the sudden change in decay rates of the GMQD that of the quantum discord occurs simultaneously if the states are the Bell diagonal states.

\section{Conclusion and discussion \label{sec:Conclusion}}

In conclusion, we have considered the dynamics of the GMQD under
decoherence.  We showed that the GMQD of a two-qubit state can be
obtained through the singular values of a special $3\times4$ matrix
whose elements are the expectation values of the Pauli matrices of the
two qubits.  With the help of the Heisenberg picture, we got the
analytic results of the geometric measure of the quantum discord for
states under three typical kinds of quantum decoherence channels. We
showed that the sudden change in decay rates of the GMQD does not
always imply that of the quantum discord, and vice versa. And at least for the Bell diagonal states, their individual sudden changes in decay rate are accordance.

In the present work, we adopt the Hilbert-Schmidt norm as the distance between two states. Besides, there exist other quantities for measuring the distance, e.g. the relative entropy~\cite{Modi2010} and the Bures distance. For the Hilbert-Schmidt distance, we show the disagreement between the GMQD and the quantum discord on reflecting the sudden changes in decay rates, so what about other kinds of the distance? Is this disagreement a property of the general geometric measure of quantum discord, or just of the geometry measure based on the Hilbert-Schmidt distance? Most recently, the phenomena of the sudden change in decay rates have been tested experimentally~\cite{JSXu2010}. 

\section{Acknowledgment}

The authors thank B.~Daki\'{c} and M. Ali for helpful
communications, and the referees for the suggestion making the present work improved.
X.~Wang is supported by NSFC with grant No.\ 10874151,
10935010, NFRPC with grant No.\ 2006CB921205; and the Fundamental Research
Funds for the Central Universities.  Z. Xi is supported by the
Superior Dissertation Foundation of Shaanxi Normal University
(S2009YB03). Z.~Sun is supported by the National Nature Science
Foundation of China with Grants No.\ 10947145; Science Foundation of
Zhejiang Province with Grant No.\ Y6090058.


\begin{thebibliography}{99}
\bibitem{Ollivier2001} H. Ollivier and W. H. Zurek,
\prl \textbf{88}, 017901 (2001).

\bibitem{Henderson2001} L. Henderson and V. Vedral, 
J. Phys. A \textbf{34}, 6899 (2001).

\bibitem{Datta2008} 
A. Datta, A. Shaji, and C. M. Caves, 
\prl \textbf{100}, 050502 (2008).

\bibitem{Lanyon2008} 
B. P. Lanyon, M. Barbieri, M. P. Almeida, and  A. G. White, 
\prl \textbf{101}, 200501 (2008).

\bibitem{Luo2008} 
S. Luo, 
\pra \textbf{77}, 042303 (2008).

\bibitem{Dillenschneider2008} 
R. Dillenschneider, 
\prb \textbf{78}, 224413 (2008).

\bibitem{Sarandy2009}
 M. S. Sarandy, 
\pra \textbf{80}, 022108 (2009).

\bibitem{Ali2010} 
M. Ali, A. R. P. Rau, and G. Alber, 
\pra, \textbf{81}, 042105 (2010).

\bibitem{Adesso2010} 
G. Adesso and A. Datta,
\prl, \textbf{105}, 030501 (2010).

\bibitem{Giorda2010} 
P. Giorda and M. G. A. Paris, 
\prl, \textbf{105}, 020503 (2010).

\bibitem{YXChen2010} 
Y.-X Chen and S.-W Li, 
\pra \textbf{81}, 032120 (2010).

\bibitem{Mazzola2010} 
L. Mazzola, J. Piilo, S. Maniscalco,
arXiv:1001.5441v2.

\bibitem{Maziero2010} 
J. Maziero, T. Werlang , F. F. Fanchini, L. C. C\'{e}leri, and R. M. Serra, 
\pra \textbf{81}, 022116 (2010).

\bibitem{Maziero2009} 
J. Maziero, L. C. C\'{e}leri, R. M. Serra, and V. Vedral, 
\pra \textbf{80}, 044102 (2009).

\bibitem{Werlang2009}
 T. Werlang, S. Souza, F. F. Fanchini, and C. J. Villas Boas, 
\pra \textbf{80}, 024103 (2009).

\bibitem{BWang2010} 
B. Wang, Z.-Y Xu, Z.-Q. Chen, and M. Feng, 
\pra \textbf{81}, 014101 (2010).

\bibitem{Fanchini2009} 
F. F. Fanchini, T. Werlang, C. A. Brasil, L. G. E. Arruda, and A. O. Caldeira, 
\pra \textbf{81}, 052107 (2010).

\bibitem{JSXu2010} 
J.-S. Xu, X.-Y. Xu, C.-F. Li, C.-J. Zhang, X.-B. Zou, and G.-C Guo, 
Nat. Commun. \textbf{1}, 7 (2010).

\bibitem{Dakic2010} 
B. Daki\'{c}, V. Vedral, and \v{C}. Brukner,
  arXiv:1004.0190v1.

\bibitem{Modi2010} K. Modi, T. Paterek, W. Son, V. Vedral, and M.
  Williamson, \prl \textbf{104}, 080501 (2010).

\bibitem{YXChen2010_2} Y.-X. Chen and Z. Yin, arXiv:1002.0176v1
  (2010).

\bibitem{Ferraro2009} A. Ferraro, L. Aolita, D. Cavalcanti,
  F. M. Cucchietti, and A. Acin, arXiv:0908.3157v3.

\bibitem{Werlang2010} T. Werlang and G. Rigolin, \pra \textbf{81},
  044101 (2010).

\bibitem{Maziero2010_2} J. Maziero, L. C. C\'{e}leri, and R. M. Serra,
  arXiv:1004.2082v1.

\bibitem{Bylicka2010} B. Bylicka and D. Chru\'{s}ci\'{n}ski,
  arXiv:1004.0434v1.

\bibitem{Datta2010} A. Datta, eprint arXiv:1003.5256v1.

\bibitem{Datta2009}
A. Datta and S. Gharibian,
\pra \textbf{79}, 042325 (2009).

\bibitem{Vedral1997} V. Vedral, M. Plenio, M. Rippin, and P. Knight,
  \prl \textbf{78}, 2275 (1997).

\bibitem{TCWei2003} T.-C. Wei and P. M. Goldbart, \pra \textbf{68},
  042307 (2003).

\bibitem{Yu2004} T. Yu and J. H. Eberly, \prl \textbf{93}, 140404
  (2004).

\bibitem{Eberly2007} J. H. Eberly and T. Yu, Science \textbf{316}, 555
  (2007).

\bibitem{Yu2009} T. Yu and J. H. Eberly, Science \textbf{323}, 598
  (2009).


\bibitem{Yu2007} T. Yu and J.H. Eberly, J. Mod. Opt. \textbf{54}, 2289
  (2007).

\bibitem{Zhou2010A} D. Zhou and R. Joynt, arXiv:1006.5474.

\bibitem{Zhou2010B} D. Zhou, G.-W. Chern, J. Fei, and R. Joynt,
  arXiv:1007.1749.

\bibitem{Koashi2004} M. Koashi and A. Winter, Phys. Rev. A
  \textbf{69}, 022309 (2004).

\bibitem{Cen2010} L.-X. Cen, X.-Q. Li, J.S. Shao, and Y.J. Yan,
  arXiv:1006.4727.

\bibitem{Groisman2005} B. Groisman, S. Popescu, and A. Winter, \pra
  \textbf{72}, 032317 (2005).

\bibitem{Schlienz1995} J. Schlienz and G. Mahler, \pra \textbf{52},
  4396 (1995).

\bibitem{Wang2010} X. Wang, A. Miranowicz, Y.-X Liu, C. P. Sun, and
  F. Nori, \pra \textbf{81}, 022106 (2010).

\bibitem{NielsenBook} M. A. Nielsen, I. L. Chuang, \emph{Quantum
    Computation and Quantum Information,} (Cambridge University Press,
  Cambridge, UK, 2000).

\bibitem{PreskillLect} J. Preskill, \emph{Lecture Notes for Physics
    219: Quantum Information and Computation} (Caltech, Pasadena, CA,
  1999), Chap. 3,
  \url{http://www.theory.caltech.edu/people/preskill/ph229/}.

\bibitem{Horodecki1996} R. Horodecki and M. Horodecki, \pra
  \textbf{54}, 1838 (1996).

\bibitem{Hill1997} S. Hill and W. K. Wootters, \prl \textbf{78}, 5022
  (1997).

\bibitem{Wootters1998} W. K. Wootters, \prl \textbf{80}, 2245 (1998).

\end{thebibliography}
\end{document}